\documentclass{mem}
\usepackage{natbib}\usepackage{txfonts}\usepackage{balance}
\usepackage{graphicx}
\usepackage[a4paper,breaklinks,dvipdfm]{hyperref}
\idline{84}{1}
\usepackage{txfonts} 

\begin{document}

\title{
Mass loading and knot formation in AGN jets by stellar winds.
}

   \subtitle{}

\author{
M. Huarte-Espinosa\inst{1}, E. G. Blackman\inst{1},
A. Hubbard\inst{2} \and A. Frank\inst{1}
          }

  \offprints{M. Huarte-Espinosa et al.}

\institute{
Department of Physics and Astronomy, University of
Rochester, Rochester NY, USA
\and
Department of Astrophysics, American Museum of Natural 
History, New York NY, USA
\email{martinhe@pas.rochester.edu}
}

\authorrunning{Huarte-Espinosa}

\titlerunning{AGN jet mass loading by stellar winds}

\abstract{
Jets from active galaxies propagate from the central black hole out
to the radio lobes on scales of hundreds of kiloparsecs. The jets
may encounter giant stars with strong stellar winds and produce
observable signatures. For strong winds and weak jets, the interaction
may truncate the jet flow during its transit via the mass loading.
For weaker jets, the interaction can produce knots in the jet.  We
present recent 3DMHD numerical simulations to model the evolution
of this jet-wind interaction and its observational consequences.
We explore (i) the relative mechanical luminosity of the radio jets
and the stellar winds (ii) the impact parameter between the jets'
axis and the stellar orbital path (iii) the relative magnetic field
strength of the jets and the stellar winds. 
}
\maketitle{}

\section{Introduction}
Powerful radio jets are often observed emerging from Active Galactic
Nuclei (AGN).  Given the stellar density at the center of
these galaxies, interactions between jets and stars may 
%
%%%alex
   %happen.
   %Jets will be mass-loaded then, and perhaps temporarily truncated,
   %by stellar winds \citep{1}. 
   occur. The stellar wind will mass-load the jets, potentially
   temporarily truncating them (Hubbard \& Blackman 2006).
%%%
%
We are carrying out numerical simulations to follow AGN jet mass
loading by stellar winds for a range of jet mechanical powers,
$L_j$, and magnetic field strengths. 
The following questions remain open: what are the observational consequences
of the jet/wind interaction?  Does the jet/wind interaction cause
radio knots as in Cen~A \citep{2}?  Do radio jet magnetic fields
affect the interaction?
\section{Model}
We use \textit{AstroBEAR2.0} \citep{3} 
to solve the equations of MHD, 
with domain: $|x|,|y| \le$2\,kpc; 0$\le z \le$6\,kpc; 
40$\times$40$\times$120~cells plus 5 AMR levels; max resolution of 
1.5\,pc. \textit{The star:} moves across the AGN jet beam at 
600\,km\,s$^{-1}$ and injects a spherical wind with 
$\dot{M}=$10$^{-4}$\,M$_{\odot}$\,yr$^{-1}$ and $v=$\,200\,km\,s$^{-1}$, 
for 10\,kyr, based on \citet{4}. We explore AGN jet mechanical
luminosities, $L_{AGN}$, of 1.5 and
0.024\,$\times$\,10$^{45}$\,erg\,s$^{-1}$
(based on \citealp{5,6}, respectively).

\begin{figure*}[t]
\begin{center}
\includegraphics[width=.5\columnwidth, bb= 100 1395 550
1470,clip=]{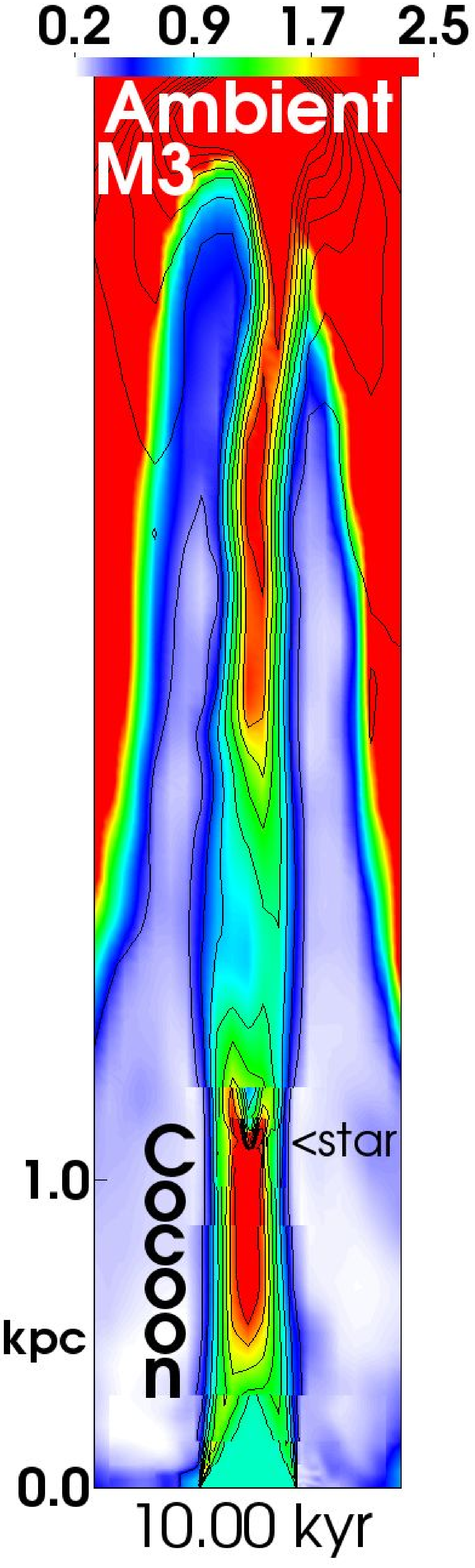}\\ 
\includegraphics[width=.403\columnwidth,bb= 110 165 485 1340,clip=]
{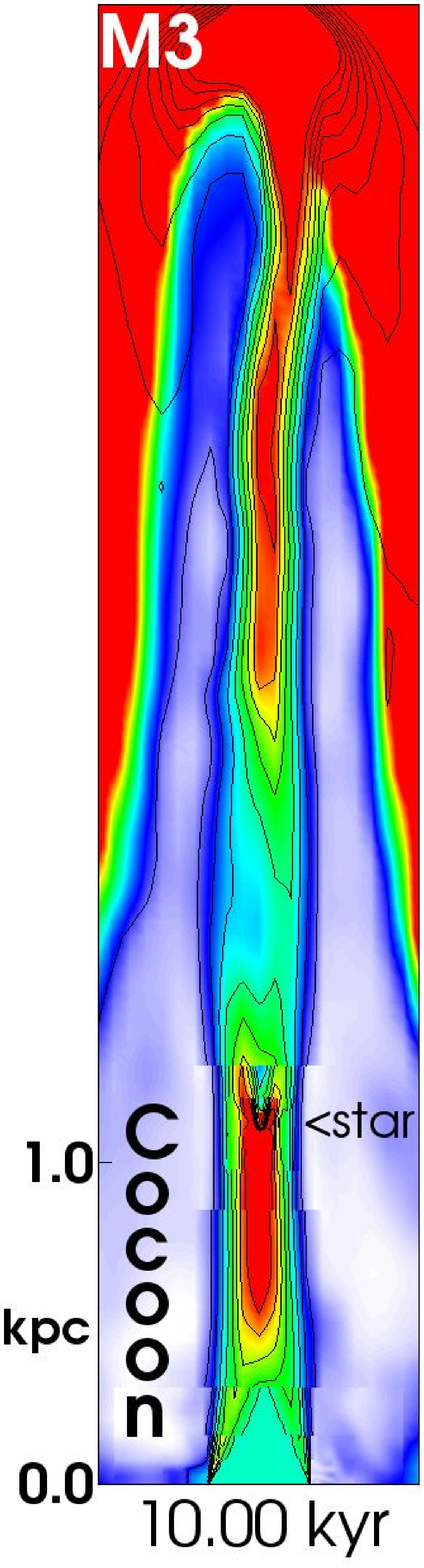}
\includegraphics[width=.29\columnwidth,bb= 220 165 485 1340,clip=]
{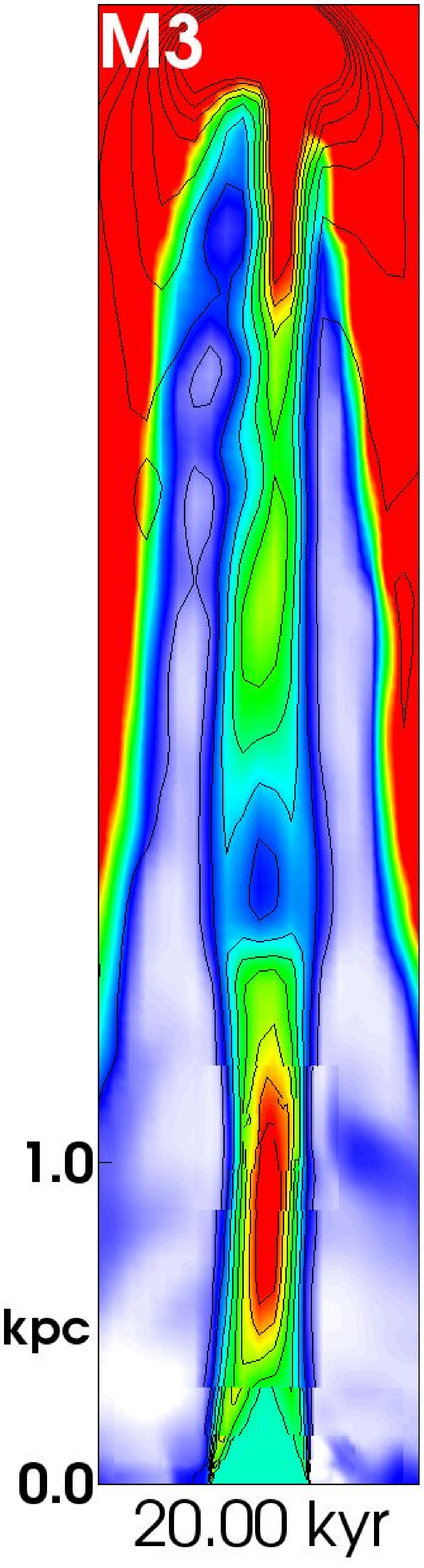}
~~~~~~~
\includegraphics[width=.29\columnwidth,bb= 220 165 485 1340,clip=]
{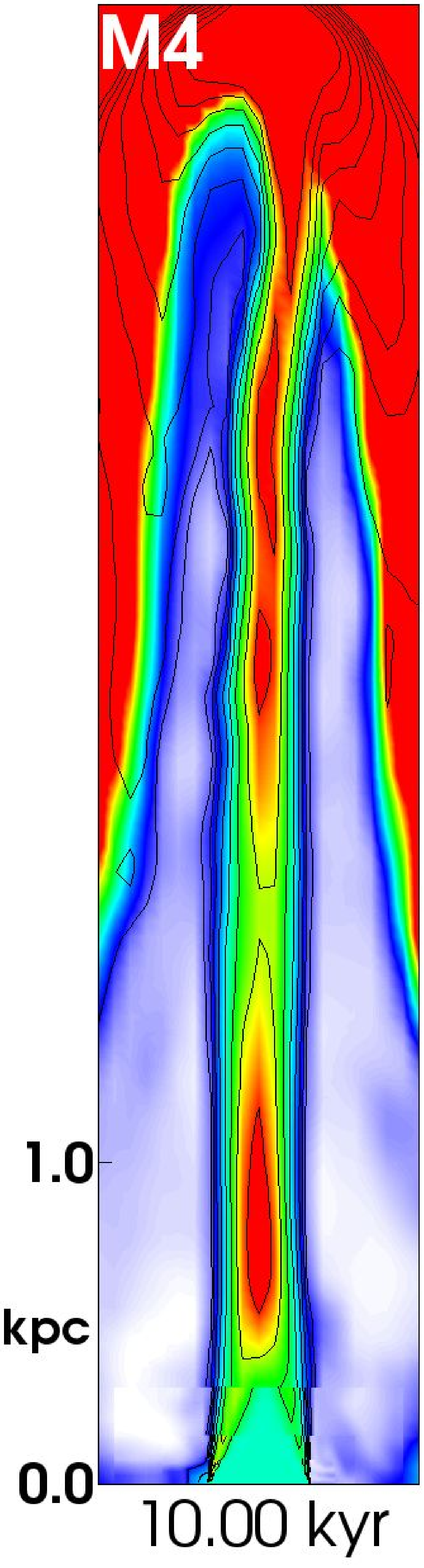}
\includegraphics[width=.29\columnwidth,bb= 220 165 485 1340,clip=]
{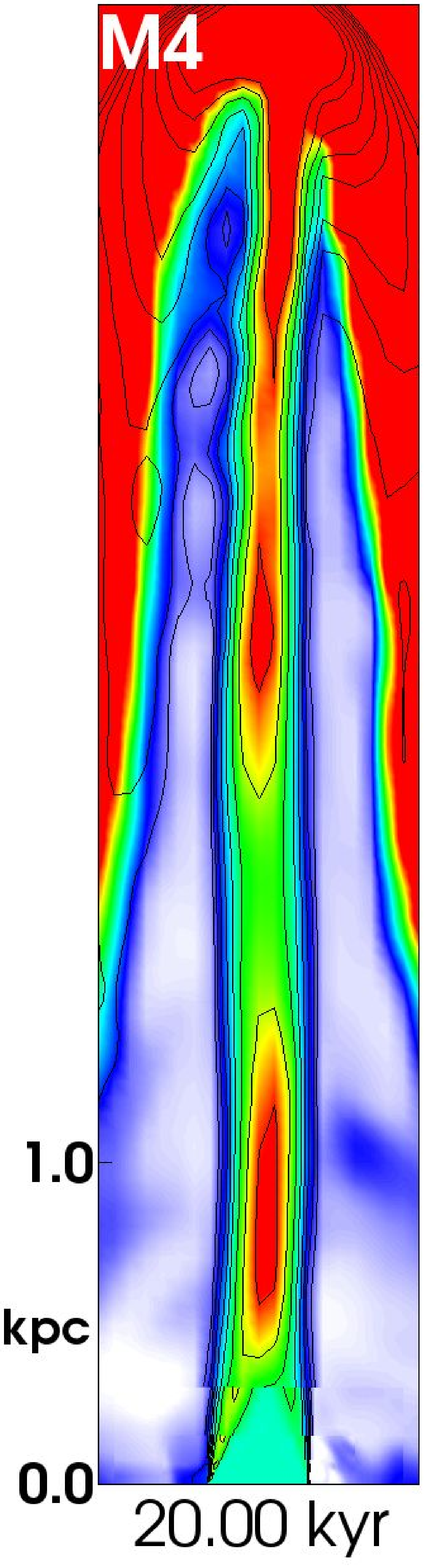}
\end{center}
\caption{\footnotesize
Density [$\times$10$^{-4}$\,part\,cm$^{-3}$]
of models with and without a wind-injecting star (left and right panels, 
respectively) for $L_{AGN}=$\,24\,$\times$\,10$^{42}$\,erg\,s$^{-1}$.
}
\label{fig}
\end{figure*}

\section{Results}
The initial jet structure in Fig. 1 is caused by fluting instabilities. 
The star injects a wind for 10\,kyr (left panel, 10\,kyr).
Mass-loading onto the AGN jet causes a density dip in the beam and
mixing of jet and stellar wind material (contours).
We see a knot in the jet's beam, which lasts for
about 30\,kyr and should have observational consequences. 
Synthetic synchrotron
emission maps will show whether the mass loading formed knot may
be distinguished from jet intrinsic structure.

\section{Conclusions}
Preliminary studies show: 
A star with $\dot{M}_{wind} =\,$10$^{-−4}$\,M$_{\odot}$\,yr$^{-−1}$
which lasts 10\,kyr is able to cause a significant density dip by
about~65\% on an AGN jet with 
$L_j=\,$2.4$\times$10$^{43}$\,erg\,s$^{-1}$.
$\bullet$ Mass loading causes mixing of jet and star material.
$\bullet$ Knots formed by stars entering the jet will propagate
along the path of the stellar orbit. This path need not be parallel
to the jet flow, thereby distinguishing such knots from those
produced by instabilities propagating along the jet. 
%Observations at different epochs should
%find that the trajectory of our model knots is different from
%that predicted in jet structure propagation models. 
$\bullet$ For models in which the entire jet volume emits, not just
the edge, the effect of mass-loading could be a dark spot.

\begin{acknowledgements}
  \scriptsize
  Financial support for this project was provided by the Space Telescope
  Science Institute grants HST-AR-11251.01-A and HST-AR-12128.01-A;
  by the National Science Foundation under award AST-0807363; by the
  Department of Energy under award DE-SC0001063; and by Cornell
  University grant 41843-7012.
\end{acknowledgements}

\bibliographystyle{aa}

\begin{thebibliography}{}
  \scriptsize

\bibitem[{Hubbard \& Blackman(2006)}]{1} 
Hubbard, A. \& Blackman, E. G., 2006, MNRAS, 371, 1717

\bibitem[Hardcastle et al.(2003)]{2} 
Hardcastle, M. et al., 2003, ApJ, 593, 169

\bibitem[Cunningham et al.(2009)]{3}
Cunningham, A. et~al., 2009, ApJS, 182, 519

\bibitem[Maeder \& Meynet(1987)]{4}
Maeder, A., \& Meynet, G. 1987, A\&A, 182, 243

\bibitem[Falcke \& Biermann(1995)]{5}
Falcke, H., \& Biermann, P.~L., 1995, AAP, 293, 665

\bibitem[Biretta, Stern \& Harris(1991)]{6}
Biretta J. A., Stern C. P., Harris D. E., 1991, AJ, 101, 1632

\end{thebibliography}

\end{document}